# The First Principle of Big Memory Systems

Yu Hua

Huazhong University of Science and Technology, csyhua@hust.edu.cn

**ABSTRACT**

In order to deliver high performance and obtain cost efficiency, Big Memory coalesces heterogeneous and distributed memory resources and exhibits salient features of large capacity, high speed and byte-addressable designs. As an important property of big memory, persistence generally needs to store temporary data in non-volatile devices and avoid data loss, which however requires extra operations and incurs system overheads. It is widely discussed if we need to guarantee persistence in the context of big memory due to the concerns about the negative impacts on system performance. From a new architecture viewpoint, persistence-related operations not only are executed off the critical path, but also simplify complex recovery mechanisms. System performance and crash consistency are further guaranteed, while avoiding frequent I/O and extra memory consumption for replica. Hence, we believe that persistence is the First Principle of Big Memory.

**BIG MEMORY**

Conventional memory designs follow Pareto principle, in which maintaining 20% hot data can meet 80% requests. The emergence of large-scale applications, such as generative AI, ChatGPT, Sora, recommendation systems, big data, and HPC systems, have fueled the demands for large-capacity and high-speed memory technologies and is changing the power-law locality. For example, LLaMA (ReGLU)-70B has to maintain 43% of neurons for 80% of total activations, which requires big memory[8]. Memory is one of the most contended hardware resources and often becomes the performance bottleneck in the I/O critical path that connects computing with storage. Due to limited memory capacity, more applications have to explore and exploit the memory beyond traditional local main memory, including empty memory on remote servers, and disaggregated memory in a memory pool. This non-local far memory can expand memory size and avoid memory stranding. Since far memory is much slower than local memory, existing systems leverage local memory as a cache by transparently swapping memory pages between local and far memory, which unfortunately causes read or write amplification.

For both local and far memory, to deliver high performance and achieve cost-efficiency, we need to carefully consider the memory tax that comes from the differential requirements of high-level applications and low-level devices. Specifically, high-level applications generally include peer-to-peer, blockchain, artificial intelligence, computability storage, etc. Different storage applications exhibit different needs and access patterns. On the other hand, the low-level devices become heterogeneous and include tapes, hard disks, PCM, ReRAM, DRAM, cache, and registers, which contain different physical properties[2]. These two-level requirements, like a hamburger, introduce the high memory tax, which means that to support program execution in memory, we need to consume extra costs for data movements and energy consumption, as well as high-security risks[1].

To alleviate high memory tax, we build big memory that not only provides large capacity but also supports fast persistence. The persistence guarantees a complete lifetime for data to be finally written into non-volatile devices via protocol buffers or checkpoints. Since most operations can be executed and completed within the big memory, we significantly reduce data movements through the memory and storage hierarchy. Transforming conventional memory to big memory leverages vertical and horizontal extensions to existing memory hierarchies. Specifically, the vertical extension allows each node to contain more data with increasing capacity and deliver high performance within flattened hierarchy, which coalesces the traditional memory and storage. Moreover, horizontally, system scales are extended to construct distributed systems. More data are aggregated together via atomic operation guarantee, which actually coalescences the central and distributed designs. Hence, by leveraging the vertical and horizontal ways, we build a big computer, as shown in Figure 1.

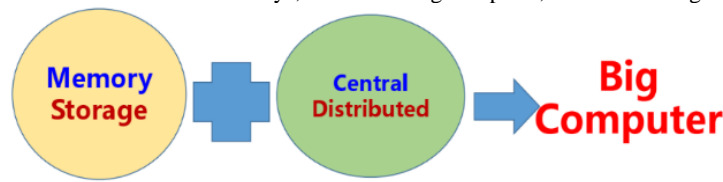

Figure 1. Big computer coalesces the memory and storage, and the central and distributed designs.

**METHODOLOGY AND IMPLEMENTATIONS**

The rationale comes from an important methodology, i.e., the computer as a network, as shown in Figure 2. In general, a computer hierarchy contains low-level storage, memory, and a high-level CPU. From the network viewpoint, the corresponding items include remote clouds, adjacent edges, and clients. In the context of a big computer, these items share some similarities. The long-latency storage seems like a remote cloud, the high-speed memory is like an adjacent edge, and the computing units are like clients that generate data. After being generated from clients/CPUs, data will be instantaneously transmitted to the close-by edges/memory, and consume a long latency to arrive at the remote clouds/storage.

The methodology mentioned can be implemented via existing devices. For example, Intel Optane DC offers about 6TB of non-volatile memory capacity, which can be considered the vertical extension. Moreover, from the horizontal view, RDMA and CXL protocols support memory and cache coherence[4]. The big computer hence achieves persistence in both vertical and horizontal extensions. A program in a big computer can be atomically executed in an end-to-end way, whether from CPU to non-volatile devices or from clients to clouds.

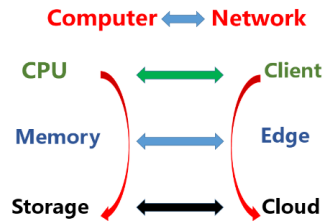

Figure 2. The "computer as a network" methodology in the context of big computers.

**UNDERSTANDING THE "BIG"**

In the context of BIG memory, we need to clarify the BIG, which is also the essential difference between big memory and traditional one. To fully understand the mentioned BIG, we first discuss other related BIG terms in the research community, as shown in Figure 3. Specifically, the difference between data and Big data is the well-known 5 Vs, including volume, velocity, variety, veracity, and value. Moreover, the difference between a model and a Big model comes from algorithms, data, and computation. Compared with chips, the Big chip needs to consider the programmable, scalable, cross-platform, highly available, and adaptive properties. The salient features behind BIG are their First Principles.

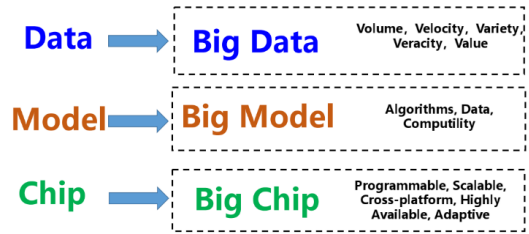

Figure 3. Understanding the "BIG" in existing terms.

We need to consider the difference between memory and big memory, and try to answer the question "What is the First Principle of big memory systems?" From the vertical view, the evolution from memory to big memory demonstrates a larger capacity and longer distance. From the horizontal view, it means the increasing complexity and scales. For example, many GPUs have been used for training and inference of large models. Horizontal scaling (e.g., distributed shared memory) temporarily maintains the intermediate results as replicas in remote volatile memory, which offers probabilistic persistence with high risks of data loss. In fact, without a persistence guarantee, numerous replicas (i.e., volatile checkpoints) occupy much memory space and incur high broadcast overheads for checkpoints and system recovery[6]. The checkpoint state sizes of typical large language models become TB-scale (e.g., 2.4TB for GPT-3), which is much larger than general 80GB GPU memory. A remote GPU hence fails to contain a whole checkpoint, which exacerbates the entire system performance.

Contemporary large-memory architectures have fundamentally reimagined data persistence, transforming it from a system constraint into a powerful performance enabler. This shift is driven by two hardware-optimized approaches: speculative persistence employs parallel RDMA replication to maintain multiple synchronized copies across nodes, enabling non-blocking durable operations, while deterministic persistence leverages CXL-connected NVM with efficient cache management (eADR) to deliver memory-tier durability without traditional storage penalties. The result is a paradigm where durability accelerates rather than hinders operations - in-memory databases achieve instant commits, machine learning systems maintain seamless checkpoints, and distributed applications preserve consistency at memory speeds. By merging the performance of volatile memory with the reliability of persistent storage, this advancement is reshaping the foundation of memory-centric computing.

## THE FIRST PRINCIPLE

We begin with a well-known debate about the Von Neumann or Non-Von Neumann architectures. The Von Neumann architecture is like hydrogen and oxygen, which work together with physical changes and exhibit some association. Furthermore, the Non-Von Neumann seems like coalesced water with chemical changes, which demonstrates the integration. Since both Neumann and Non-Von Neumann architectures need to persist data, they can be unified in the context of persistence. We hence argue that persistence is the First Principle of big memory systems, as shown in Figure 4.

In practice, the memory/storage hierarchy in the Von Neumann architecture is a multi-level subset model, which includes tape, disks, SSD, PCM, DRAM, L1/L2/L3 caches, and registers. In the subset model, the devices in the lower levels contain the data in the higher ones. All data need to be written into persistent devices. On the other hand, for Non-Von Neumann architecture, persistence still plays an important role. For example, a typical workflow in the memristors needs to compute the input vector in the matrix and the result is a new vector that is volatile. By using extra operations, the new vector can be finally persisted in the non-volatile devices. Hence, both Von Neumann and Non-Von Neumann architectures follow the persistence principle.

There exist some non-persistence operations in conventional systems, e.g., replicating data into the DRAMs across servers or writing data in a batch through a cache/memory/storage hierarchy. These operations often cause misleading understanding that persistence is harmful to high performance, which is not. To avoid the high overheads of re-reading or re-computing the lost data, systems persist data but suffer from the huge gap between high-speed memory and low-speed disks. Many efforts have to be put into maintaining data in a volatile manner, which is vulnerable to just one system crash or failure. This awkward situation can be efficiently addressed using the big memory that offers large-size non-volatility with a shortened critical path. A case in point is eADR[1] which offers persistence for in-cache data and guarantees to flush them into non-volatile devices in case of system crash or failure. Hence, the number of flushing data is significantly reduced even during normal system execution. Systems thus deliver high performance with the aid of persistence.

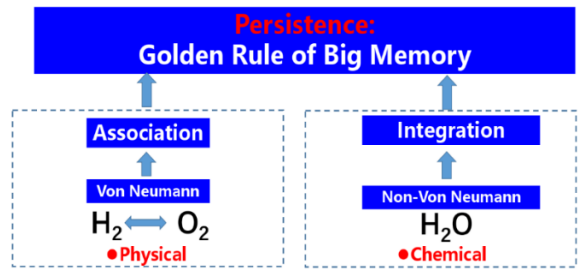

Figure 4. Persistence is the First Principle of big memory systems.

Persistence needs to be guaranteed by storing states and data in non-volatile storage devices. A program hence transfers states and data to and from storage devices via I/O bus, which is supported by leveraging specific instructions in a program. Persistence is important to deal with system failure and power down without data loss. These data can be accessed via memory instructions or APIs.

In general, operating systems run on fast, byte-addressable, volatile main memory and slow, block-interface, non-volatile storage. To store data persistently, data have to frequently and expensively move back and forth between the two-tiered memory-storage hierarchy. The separated hierarchy often results in consistency bugs and performance decrease of data swapping.

Due to the salient features of DRAM-like performance and disk-like non-volatility, persistent memory offers a flat-addressing device, which allows consistent states in memory and flushes them to storage in batch[5]. This reduces unnecessary data movements for persistence and avoids bus snooping-like security issues. The flat addressing mitigates the complicated and error-prone data persistence while supporting near-instantaneous recovery and direct manipulation of persistent data. Moreover, although Intel eADR can guarantee the persistence of data in multi-level CPU caches, the data stored in CPU registers may still be lost upon power failures, since the power capacitors cannot flush them from registers to non-volatile devices. Hence, checkpointing techniques have been further used to guarantee the persistence of consistent full-stack states.

## FULL-STACK AND MOVING PERSISTENCE

As shown in Figure 5, in the entire memory hierarchy, we achieve full-stack persistence. The persistent memory naturally offers data persistence due to physical property. ADR (Asynchronous DRAM Refresh) allows the data in the memory controller to be flushed into non-volatile devices. eADR (extended ADR) further guarantees the data in multi-level caches to be persisted into non-volatile devices. In fact, the register can also guarantee the persistence via expensive and frequent checkpoints[7]. The persistent boundary thus moves from the traditional boundary between SSD and DRAM to register and CPU. Therefore, existing memory hierarchy supports full-stack persistence starting from the CPU computation units, thus achieving a larger capacity.

Full-stack persistence requires support from non-volatile devices and periodic checkpointing to facilitate fast recovery from system crashes or power failures. Although this hierarchy supports full-stack persistence, the implementations are different. For persistent memory, persistence can be naturally obtained due to its physical features. However, the memory controller and multi-level caches, which are volatile, require extra power to flush data into non-volatile devices, called moving persistence. Since the data moves via the I/O bus, there exist security issues due to bus snooping. These problems can be addressed by efficiently coordinating the hierarchy, capacity, and operations for high performance and strong security, thus achieving a suitable tradeoff between performance and security.

For large-scale distributed systems, we conventionally leverage IP protocol to connect the hard disks in multiple nodes. RDMA is further used to aggregate memory resources together. CXL can achieve the cache coherence[3]. These designs allow the processing boundary to move up and more operations can be completed in the higher levels. Since data must be written into non-volatile devices, the end-to-end critical path can be significantly shortened. We hence achieve the savings of data movements and energy consumption, as well as alleviating bus conflicts and side-channel attacks.

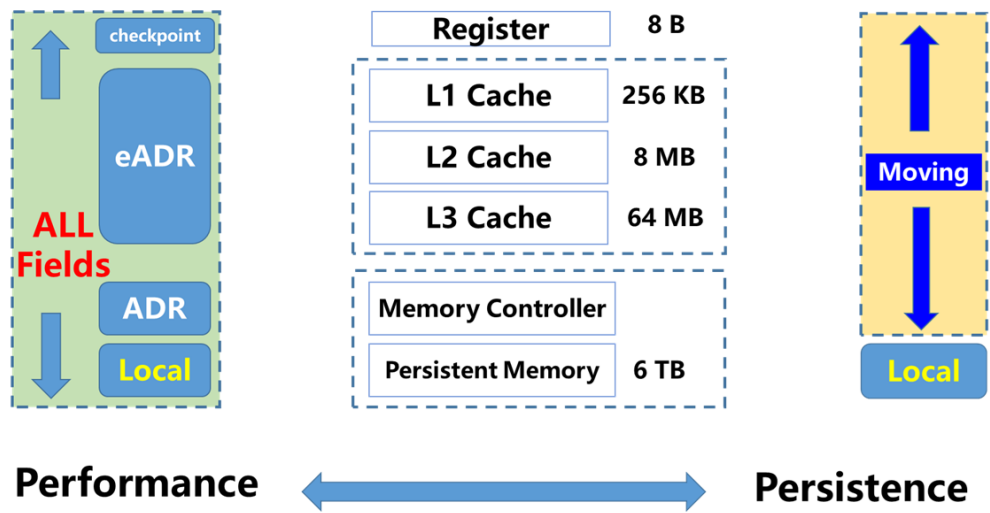

Figure 5. The full-stack and moving persistence.

**SHAPING THE FUTURE**

To guarantee the persistence of big memory, multi-tier memory systems rethink the need for data tiering, or moving the data accessed by the program to a memory pool for matching the access patterns. Specifically, the memory hierarchy is adding more tiers to narrow their performance gap and becoming more heterogeneous to meet the performance, cost, and capacity requirements of applications. Multi-tier memory techniques, such as multi-socket Non-Uniform Memory Access (NUMA), High Bandwidth Memory (HBM), and Compute Express Link (CXL), construct scalable and cost-effective memory systems, which exceed conventional two layers. CXL can seamlessly integrate persistent memory, graphics processing units, and accelerators, which allow direct memory access to heterogeneous devices. Sophisticated checkpointing techniques offer efficient persistence and update model parameters and embeddings across training batches, achieving significant boosts in training performance and energy savings.

Modern networking technologies like RDMA and CXL are erasing the boundaries between memory and storage, turning distributed resources into a seamless, low-latency memory fabric. CXL pools memory across servers, enabling elastic allocation at near-DRAM speeds, while RDMA skips software overhead with direct peer-to-peer data transfers. Instead of slow block storage, systems now operate on byte-addressable memory—where persistence is a property, not a separate hardware tier. This evolution makes traditional CPU caches, DRAM, and disk hierarchies inefficient, replacing them with a scalable network-spanning memory system that grows in performance and capacity as the infrastructure expands.

In the context of big memory, CXL allows CPUs to directly access PCIe (Peripheral Component Interconnect Express) devices through load/store instructions, and effectively improves system scalability, since more PCIe devices can be connected across switches. To provide a cost-effective way to enhance memory bandwidth and capacity, CXL introduces caching coherency and memory semantics for devices directly connected to the host. In this way, we can handle fine-grained heterogeneous processing of the shared data structures between CPUs and accelerators. CXL can transparently expand the memory capacity, achieve elastic memory resource sharing, and allow applications to access remote memory via pure load/store instructions. These coalesced designs will offer the great potential of guaranteeing persistence and improving memory scalability and hardware utilization.